\documentclass[utf8]{frontiersFPHY} 

\usepackage{url,hyperref,lineno,microtype,subcaption}
\usepackage[onehalfspacing]{setspace}

\def\keyFont{\fontsize{8}{11}\helveticabold }
\def\firstAuthorLast{Moya and Navarro} 
\def\Authors{Pablo S. Moya\,$^{1,*}$ and Roberto E. Navarro\,$^{2,*}$}
\def\Address{$^{1}$Departamento de F\'isica, Facultad de Ciencias, Universidad de Chile, Chile \\
$^{2}$Departamento de F\'isica, Facultad de Ciencias F\'isicas
  y Matem\'aticas, Universidad de Concepci\'on, Chile}
\def\corrAuthor{Pablo S. Moya}
\def\corrEmail{pablo.moya@uchile.cl}

\begin{document}
\onecolumn
\firstpage{1}

\title[Quasilinear Turbulence in Space Plasmas]{Effects of the Background Turbulence on the Relaxation of Ion Temperature Anisotropy in Space Plasmas} 

\author[\firstAuthorLast ]{\Authors} 
\address{} 
\correspondance{} 

\extraAuth{Roberto E. Navarro\\roberto.navarro@udec.cl}

\maketitle

\begin{abstract}
Turbulence in space plasmas usually exhibits two regimes separated by a spectral break that divides the so called inertial and kinetic ranges. Large scale magnetic fluctuations are dominated by non-linear MHD wave-wave interactions following a -5/3 or -2 slope power-law spectrum. After the break, at scales in which kinetic effects take place, the magnetic spectrum follows a steeper power-law $k^{-\alpha}$ shape given by a spectral index $\alpha > 5/3$. Despite its ubiquitousness, the possible effects of a turbulent background spectrum in the quasilinear relaxation of solar wind temperatures are usually not considered. In this work, a quasilinear kinetic theory is used to study the evolution of the proton temperatures in an initially turbulent collisionless plasma composed by cold electrons and bi-Maxwellian protons, in which electromagnetic waves propagate along a background magnetic field. Four wave spectrum shapes are compared with different levels of wave intensity. We show that a sufficient turbulent magnetic power can drive stable protons to transverse heating, resulting in an increase in the temperature anisotropy and the reduction of the parallel proton beta. Thus, stable proton velocity distribution can evolve in such a way as to develop kinetic instabilities. This may explain why the constituents of the solar wind can be observed far from thermodynamic equilibrium and near the instability thresholds.

\tiny
 \keyFont{ \section{Keywords:} Space plasma physics, turbulence, ion-cyclotron waves, temperature anisotropy instability, quasilinear theory} 
\end{abstract}

\section{Introduction}

In many space environments the media is filled by a weakly collisional plasma. Although Coulomb collisions represent an efficient mechanism for relaxing plasma parcels towards a thermodynamic equilibrium state in which the particle Velocity Distribution Functions (VDFs) achieve a Maxwellian
profile~\citep{spitzer1962,marsch1983}, when collisions are scarce Coulomb scattering becomes ineffective in establishing equilibrium. Subsequently, kinetic collisionless processes may dominate the dynamics of the system and be responsible for many of the observed macroscopic and microscopic properties of the plasma.
Under these conditions the plasma VDF usually develops non-Maxwellian characteristics that can provide the necessary free energy to excite micro-instabilities that subsequently can induce changes on the macroscopic properties of the plasma~\citep{Marsch1982,gary1993,Astudillo1996,Hellinger2006,Lopez2020}. Among the fundamental problems of plasma physics belongs the understanding of the excitation and relaxation processes of these poorly collisional plasmas and the resultant state of nearly equipartition energy density between plasma particles and electromagnetic turbulence~\citep{klimontovich1997}. In particular, these processes play an important role in space plasma environments such as the solar wind~\citep{kasper2002,marsch2006,bale2009,bruno2013} and the Earth's magnetosphere~\citep{heinemann1999,matsumoto2006,espinoza2018},
specially at kinetic scales~\citep{gary1993,moya2015,Narita2020}.

It is well known that in space plasmas, turbulence usually exhibits two distinct regimes separated by a spectral break dividing the fluctuations power spectrum. In the case of magnetic field fluctuations these two regimes are known as the inertial MHD range (at larger scales) and the kinetic range (at ionic and sub-ionic scales)~\cite{Sahraoui2010,bruno2013}. Large scale magnetic fluctuations are dominated by MHD non-linear wave-wave interactions following a -5/3 or -2 slope power-law spectrum~\citep{Horbury2008}. After the break the spectrum follows a steeper power-law $k^{-\alpha}$ shape given by a spectral index $\alpha > 5/3$, as it has been shown from observations~\citep{Chen2016}, 2.5D simulations~\citep{Gonzalez2019}, and full 3D simulations \citep{Cerri2019}. The break is related with the scales in which kinetic effects such as wave-particle interactions~\citep{Howes2008,maneva2015,sulem2015} take place, or ion-scale current sheets are disrupted due to the onset magnetic reconnection~\citep{Cerri2017,Franci2017,Loureiro2017,mallet2017,Papini2019}. Depending on the local plasma conditions the break can coincide with the ion inertial length $\lambda_i$, or the ion gyroradius $\rho_i = \sqrt{\beta} \lambda_i$, where $\beta$ is the plasma beta parameter, as observed in space plasmas~\citep{Chen2014,Wang2018} and numerical simulations~\citep{Franci2016}. In particular, for small values of the proton plasma beta, the scale of the break seems to be related to $\lambda_i$; while for larger values of beta to $\rho_i$. Also, different plasma environments can exhibit different spectral indices. For example, considering Van Allen Probes observations~\citet{Gamayunov2015} have found $\alpha\sim 2$ in the inner magnetosphere, and~\citet{Chaston2014,moya2015} observed $\alpha\sim4$ associated with Kinetic Alfv\'en Waves turbulent spectra measured during geomagnetic storms in the Ring Current region. 
Similarly, \citet{Alexandrova2008} found $\alpha\sim7/3$ using Cluster data, and~\citet{Goldstein2015} observed $\alpha\sim4.2$ considering Wind observations in the case of the solar wind at 1 AU from the Sun. In addition, using data from the recent Parker Solar Probe mission~\citet{Franci2020} have found $\alpha\sim 7/2$ during the first perihelion of the spacecraft, at about 36 solar radii from the Sun. Further, all these results are consistent with kinetic scale simulations~\citep{Franci2015,Cerri2019,Franci2020}. Besides the recent progress, how the turbulent energy is dissipated in all these almost collisionless plasma systems is still under debate and corresponds to one of the outstanding open questions in space plasma physics~\citep{Chen2016,Matthaeus2020}.  

In a magnetized plasma such as the solar wind or the Earth's magnetosphere, one of the most typical deviations from the Maxwellian equilibrium is the bi-Maxwellian distribution, i.e. a composed Maxwellian VDF that exhibits different thermal spreads (different temperatures) in the directions along and perpendicular to the background magnetic field. These distributions are susceptible to temperature anisotropy driven micro-instabilities that can effectively reduce the anisotropy and relax the plasma towards more isotropic states. However, in the absence of enough collisions, these instabilities are usually not able to lead the system fully into thermodynamic equilibrium, and the plasma allows a certain level of anisotropy up to the so called kinetic instability thresholds~\citep{gary1993,vinas2015}. From the theoretical kinetic plasma physics point of view, on the basis of the linear and quasilinear theory approximations of the dynamics of the plasma, it is possible to predict the thresholds in the temperature anisotropy and plasma beta parameter space that separate the stable and unstable regimes, and how the plasma evolves towards such states. These models are useful to study the generation and first saturation of the electromagnetic energy at the expense of the free energy carried by the plasma. To do so, in general quasilinear calculations consider initial conditions with a small level of magnetic field energy that grows as the temperature anisotropy relaxes. A comprehensive review of linear and quasilinear analysis of these instabilities considering a bi-Maxwellian model can be found in~\citet{Yoon2017} and references therein.

Since the first studies by~\citet{weibel} and~\citet{sagdeev1960}, the research about temperature anisotropy driven modes and the stability of the plasma have been widely studied in the last decades, and represent an important topic for space plasmas physics~\citep{kennel1966b,vinas1984,brinca,yoon,moya2014}.
Predictions based on a bi-Maxwellian description of the plasma are qualitatively in good agreement with observations of solar wind protons (see e.g.~\citet{Hellinger2006,bale2009}) and electrons (see e.g.~\citet{hellinger2014,adrian2016}). However, as mentioned, turbulence is ubiquitous in space environments and all these relaxation processes should occur in the presence of a background turbulent magnetic spectrum. To the best of our knowledge, only a few quasilinear studies such as~\citet{Moya2012} or~\citet{moya2014} have considered a background spectrum but nonetheless a study focused on the possible effects of a magnetic field background spectrum is yet to be done. Here we perform such systematic study by computing the quasilinear relaxation of the ion-cyclotron temperature anisotropy instability, considering different choices of the initial level of the magnetic field fluctuations, and the shape of the spectrum. We analyze their effect on the relaxation of the instability and the time evolution of the macroscopic properties of the plasma that are involved. 

Several studies of the solar wind electromagnetic turbulence near the spectral break as a function of wave number $|k| = \sqrt{k^2_\perp + k^2_\parallel}$, where $k_\perp$ and $k_\parallel$ are the wave vector components parallel and perpendicular to the magnetic field, respectively, have shown that the fluctuation spectrum is anisotropic and that the power spectrum sometimes is greater at quasi-perpendicular propagation $k_\perp \gg k_\parallel$ (see e.g. \citet{Dasso2005,Horbury2005}) than at quasi-parallel propagation $k_\perp \ll  k_\parallel$. However, for small plasma beta ($\beta<1$) as in this study, the compressibility of the magnetic fluctuations is small in the solar wind at 1 AU (see e.g. \citet{bale2009}), which is consistent with propagation of Alfv\'en ion-cyclotron waves. Therefore, as a first approximation we consider the fluctuations to be magnetically non-compressive and propagating strictly along the background magnetic field ($k=k_\parallel$), and also have and focused on small $\beta$ values. In the next section we present the details of our quasilinear model, and then, in Sections~\ref{sec:Results1} and \ref{sec:Results2} we present and discuss all our numerical results. Finally, in the last section we summarize our findings and present the main conclusions of our work.

\section{Quasilinear temperature evolution}
\label{sec:disprel}
We consider a magnetized plasma composed of bi-Maxwellian protons and cold electrons. The kinetic dispersion relation of left-handed circularly polarized waves, propagating along a background magnetic field $\vec{B}_0$ is given by~\citep{Moya2012,Gomberoff2004,Navarro2020}
\begin{equation}
    \frac{v_A^2 k_\parallel^2}{\Omega_p^2} = A + (A\xi^{-} + \xi)Z(\xi^{-}) - \frac{\omega_k}{\Omega_p} \,, \label{eq:disprel}
\end{equation}
where $\omega_k = \omega + i\gamma$ is the complex frequency that depends on the wavenumber $k_\parallel$; $v_A=B_0/\sqrt{4\pi n_p m_p}$ is the Alfvén speed, with $n_p$ and $m_p$ the density and mass of protons, respectively; $\Omega_p=eB_0/m_pc$ is the proton gyrofrequency with $c$ the speed of light; $A=T_\perp/T_\parallel-1$ where $T_\perp/T_\parallel$ is the temperature anisotropy; $T_\perp$ and $T_\parallel$ are the proton temperatures perpendicular and parallel with respect to $\vec{B}_0$, respectively; $\xi=\omega_k/k_\parallel u_\parallel$ and $\xi^{-}=(\omega_k-\Omega_p)/k_\parallel u_\parallel$ are resonance factors~\citep{Gary1985}; $u_\parallel=\sqrt{2k_BT_\parallel/m_p}$ is the parallel proton thermal speed, and $k_B$ the Boltzmann constant. $Z(\xi)$ is the plasma dispersion function~\citep{zeta}, which is calculated numerically with the Faddeeva function provided by scipy. We also define the parallel proton $\beta_\parallel=u_\parallel^2/v_A^2$. In Eq.~\eqref{eq:disprel}, we have assumed charge neutrality (i.e. zero net charge such that the electron density is $n_e=n_p$), and $v_A/c\ll1$. Numerical roots of Eq.~\eqref{eq:disprel} are calculated through the Muller's method~\cite{Muller1956} using our own Python code. The dispersion relation Eq.~\eqref{eq:disprel} supports an infinite number of solutions for $\omega_k$ for each value of $k_\parallel$, most of them being sound-like heavily damped modes with frequencies above and below the proton gyrofrequency~\cite{Astudillo1996,Navarro2014}. Here, we focus on the quasilinear evolution of the plasma due to Alfvén-Cyclotron Wave (ACW) instabilities (for more details about this instability in the context of space plasmas see e.g.~\citet{Moya2011,jian2016,wicks2016,Yoon2017} and references therein).
\begin{figure}
  \centering
  \includegraphics[width=\linewidth]{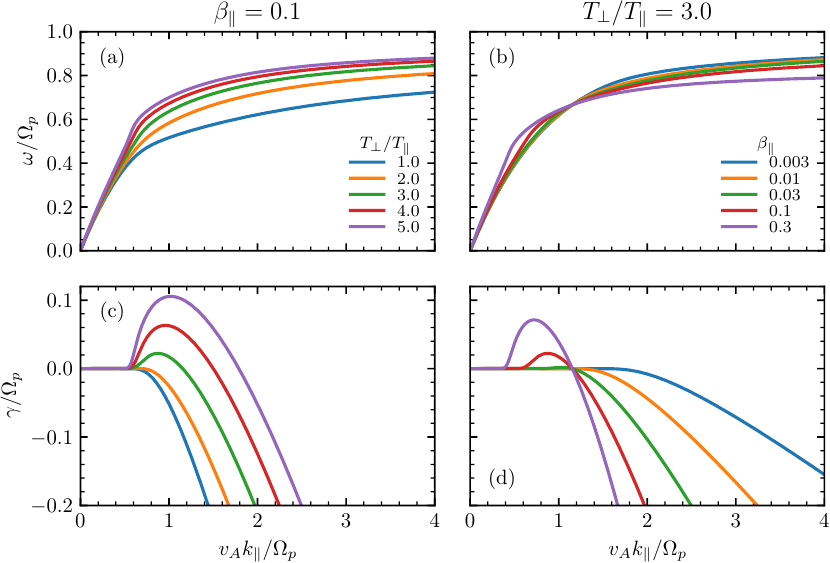}
  \caption{(top) Frequency and (bottom) growth/damping rate of Alfvén-cyclotron waves as a function of wavenumber, as calculated from the dispersion relation Eq.~\eqref{eq:disprel}. (left) $\beta_\parallel=0.1$ and different values of $T_\perp/T_\parallel$. (right) $T_\perp/T_\parallel=3$ and different values of  $\beta_\parallel$. Plots for $k_\parallel<0$ can be obtained through the parity condition $\omega_{-k}=-\omega_k^\ast$.}
  \label{fig:disprel}
\end{figure}

Figure~\ref{fig:disprel}(a) shows the real part of the ACW complex frequency for $\beta_\parallel=0.1$ and several values of the temperature anisotropy $T_\perp/T_\parallel$. Similarly,  Fig.~\ref{fig:disprel}(b) shows the effects of $\beta_\parallel$ on the ACW real frequency at a fixed anisotropy $T_\perp/T_\parallel=3$.
In all cases with $\beta_\parallel<0.1$, the real part of the frequency seems to approach asymptotically to $\omega=\Omega_p$ at large wavenumbers. This description is very similar to the solutions of the dispersion relation Eq.~\eqref{eq:disprel} in the cold-plasma approximation~\citep{krall,stix1992}. However, for $T_\perp/T_\parallel>1$ [Fig.~\ref{fig:disprel}(a)] or $\beta_\parallel>0.01$ [Fig.~\ref{fig:disprel}(b)], the frequency curve deviates from the cold-plasma approximation for wavelengths around the proton inertial length $v_A/\Omega_p$. 

Kinetic effects can damp ACWs of large wavenumbers even at low beta, and large temperature anisotropies can drive part of the wave spectrum unstable. Figures~\ref{fig:disprel}(c) and (d) show the imaginary part of the complex frequency for the same parameters as in Figs.~\ref{fig:disprel}(a) and (b), respectively. The wave is damped if its frequency satisfies ${\rm Im}(\omega_k) = \gamma <0$, or it is unstable if $\gamma>0$. Figure~\ref{fig:disprel}(c) shows that even a small value of $\beta_\parallel=0.1$ allows the growth of a kinetic instability when $T_\perp/T_\parallel\gtrsim2$ in a small range of wavenumbers $0.6\lesssim v_A k_\parallel/\Omega_p \lesssim 2$. The maximum value of $\gamma$ increases, and its wavenumber also increases, as the anisotropy rises above $T_\perp/T_\parallel\gtrsim2$. On the other hand, for $v_Ak_\parallel/\Omega_p\gtrsim2$ and $\beta_\parallel=0.1$ the ACW is always damped, with damping rate $\gamma$ decreasing almost linearly with $k_\parallel$. The wave is also marginally stable ($\gamma\approx0$) for long wavelengths compared with the ion inertial length ($v_A k_\parallel /\Omega_p \lesssim 0.5$).

Notice in Fig. 1(d) that the ACW is marginally stable ($\gamma=0)$ at a fixed wavenumber value $v_A k_\parallel/\Omega_p\sim 1.15$ and $T_\perp/T_\parallel=3$ independently of the value of $\beta_\parallel$. It can be shown from Eq.~\eqref{eq:disprel} that this happens at $(v_Ak_\parallel/\Omega_p)^2=(R-1)^2/R$~\cite{yoon2012b}. Thus, for a lower value of the temperature anisotropy the waves are marginally stable at lower wavenumbers, as seen in Fig.~\ref{fig:disprel}(c), and the damping becomes stronger as the anisotropy approaches $T_\perp/T_\parallel=1$. Also, the instability decreases both with lower $\beta_\parallel$ and lower $T_\perp/T_\parallel$. It is important to mention that a semi-cold approximation of the plasma ($\xi^{-}\gg 1$) fails to describe these properties, making it inappropriate for the quasilinear evolution of the plasma temperature.   

The quasilinear moments approximation assumes that the macroscopic parameters of the plasma evolve adiabatically, thus $\omega_k=\omega_k(t)$ solves the dispersion relation Eq.~\eqref{eq:disprel} instantaneously at all times. The quasilinear evolution of the perpendicular and parallel thermal speeds are given by~\cite{Moya2011,Moya2012}
\begin{align}
  \frac{\partial u_\perp^2}{\partial t} 
  &= 
  -{\rm Im}\,\frac{4}{L} \frac{e^2}{m_p^2}  \int_{-\infty}^{\infty}
  dk_\parallel \frac{|B_k|^2}{c^2k_\parallel^2} \left[ (2 i \gamma - \Omega_p) \left(\frac{v_A^2 k_\parallel^2}{\Omega_p^2}+\frac{\omega_k}{\Omega_p}\right) + \omega_k
  \right] \,, \label{eq:uperp}
  \\
  \frac{\partial u_\parallel^2}{\partial t} 
  &= 
  {\rm Im}\, \frac{8}{L} \frac{e^2}{m_p^2}  \int_{- \infty}^{\infty} dk_\parallel \frac{|B_k|^2}{c^2 k_\parallel^2} \left[
  (\omega_k - \Omega_p) \left(\frac{v_A^2 k_\parallel^2}{\Omega_p^2}+\frac{\omega_k}{\Omega_p}\right) + \omega_k\right] \,. \label{eq:uz}
\end{align}
where $L$ is the characteristic length of the plasma, and $|B_k|^2$ is the spectral wave energy satisfying 
\begin{equation}
\frac{\partial |B_k|^2}{\partial t}= 2\gamma(t) |B_k|^2, \label{eq:dbkdt}
\end{equation}
such that Eqs.~\eqref{eq:disprel}-\eqref{eq:dbkdt} form a closed system to address the quasilinear evolution of the ACW instability. The quasilinear approach summarized in Eqs.~\eqref{eq:disprel}-\eqref{eq:dbkdt} is a widely used theoretical approach to study non-linear effects in the evolution of plasma waves as they interact with the media. Comparisons between quasilinear solutions and hybrid or particle-in-cell simulations~\citep{seough2014,seough2015,Yoon2017} have shown that the approach is valid (theoretical and numerical results are in relatively good agreement) when the amplitude of the waves is finite but relatively small, especially for resonant instabilities (such as the ACW instability). Moreover, comparisons have also shown that the agreement between simulations and quasi-linear models is remarkable during the exponential growth of the instability (see e.g. \citet{Yoon2017}). Thus, for our calculations we have restricted the initial magnetic energy to $W_B(0)=\int d{k_\parallel}|B_k(0)|^2/B_0^2\le0.1$ (corresponding to $B_k/B_0\lesssim0.1$ for a uniform spectrum), and have followed the quasilinear time evolution up to $\Omega_pt=150$ ensuring that the plasma reaches a stationary state.

In the next sections we explore the effects of the $B_k$ spectrum on the relaxation of the proton anisotropy.

\section{Numerical Results. The effect of a background spectrum}
\label{sec:Results1}

In order to solve numerically the system of differential equations given by Eqs.~\eqref{eq:disprel}--\eqref{eq:dbkdt} we use a fourth order Runge-Kutta method. In this section, for academic purposes, we consider three distinct shapes for the magnetic spectrum $|B_k|^2$ to illustrate the effects that different initial magnetic field background spectra can produce on the quasilinear evolution of the
macroscopic parameters of the plasma. The three initial background spectra considered here are a uniform noise $|B_k(0)|^2=A$, a Gaussian spectrum
\begin{align}
    |B_k(0)|^2 = A\, e^{-(v_A {k_\parallel}/\Omega_p)^2}\,,
\end{align}
and a Lorentzian spectrum
\begin{align}
    |B_k(0)|^2 = \frac{A}{1+(v_A {k_\parallel}/\Omega_p)^\alpha}\,,
\end{align}
where the normalization constant $A$ is adjusted depending on the initial total magnetic energy $W_B(0)$, with the integral calculated in the range $10^{-3}<v_A{k_\parallel}/\Omega_p<8$. The large-wavenumber tails $v_A{k_\parallel}/\Omega_p>8$ of the spectrum shapes considered here do not contribute to the quasilinear plasma evolution. For these ${k_\parallel}$ values the waves are heavily damped as Fig.~\ref{fig:disprel} shows for $\beta_\parallel>0.003$. Thus, if energy is stored at those ${k_\parallel}$ values, they are quickly transferred to the particles until the wave energy is depleted. For greater values of ${k_\parallel}$, this process is faster. Therefore, most of the quasilinear evolution at late stages will be carried by energy transfer around $v_A{k_\parallel}/\Omega_p=1$ where the wave is marginally stable and an instability is likely to appear.

\begin{figure}[ht!]
  \centering
  \includegraphics[width=\textwidth]{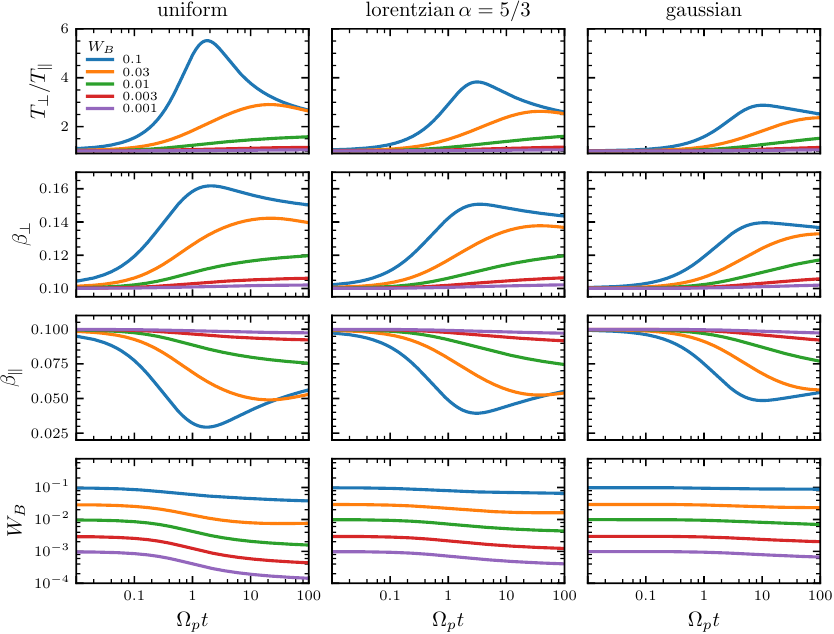}
  \caption{Quasilinear evolution of (upper row) the temperature anisotropy, (second row) perpendicular $\beta_\perp$, (third row) parallel $\beta_\parallel$, and (lower row) total magnetic energy. Initial conditions for all cases were chosen as $\beta_\parallel(0)=0.1$, $T_\perp(0)/T_\parallel(0)=1$, and different values of the initial magnetic energy: (blue) $W_B(0)=0.001$, (orange) $W_B(0)=0.003$, (green) $W_B(0)=0.01$, (red) $W_B(0)=0.03$, and (purple) $W_B(0)=0.1$. Each column shows (left) a uniform, (middle) $\alpha=5/3$ Lorentzian, and (right) Gaussian initial background spectrum. Time and $W_B$ axes are in logarithmic scale.}
  \label{fig:quasilinear}
\end{figure}

Figure~\ref{fig:quasilinear} shows the quasilinear time evolution of the temperature anisotropy, the perpendicular $\beta_\perp$ and parallel $\beta_\parallel$, and the total magnetic energy $W_B$. The initial conditions are chosen as $T_\perp(0)/T_\parallel(0)=1$ and $\beta_\parallel=0.1$, parameters that are close to the most observed values in the solar wind at 1 AU in which quasi-parallel propagation seems to be dominant~\citep{bale2009}. For every magnetic shape spectrum, we also compare the effects of different values of the initial level of magnetic fluctuations $W_B(0)=0.003$ through $0.1$. According to linear theory, the plasma is stable for the chosen initial parameters. In fact, for $\beta_\parallel=0.1$, an isotropic velocity distribution $T_\perp/T_\parallel=1$ has no free energy to excite an instability [see blue line in Fig.~\ref{fig:disprel}(c)]. Thus, we should expect that the temperatures will remain almost constant in time~\cite{Navarro2014}. However, an striking feature for all the spectrum shapes, is that the anisotropy can grow in time if a sufficient level of magnetic energy is provided. 

For an uniform spectrum of total level $W_B(0)=0.1$ (blue lines in Fig.~\ref{fig:quasilinear}), the anisotropy can grow up to high values $T_\perp/T_\parallel\simeq5$ in a small time frame. This results in a sharp increase in the perpendicular beta from  $\beta_\perp=0.1$ to $\approx0.16$, and consistently a rapid fall in the parallel $\beta_\parallel=0.1$ towards $\approx0.03$. Afterwards, the anisotropy decreases while $\beta_\parallel$ rises, both steadily, towards a quasi-stationary state around $T_\perp/T_\parallel\simeq3$ and $\beta_\parallel\approx0.6$. We note that this anisotropy growth is not as explosive for a Lorentzian (with $\alpha=5/3$, right column in Fig.~\ref{fig:quasilinear}) and a Gaussian spectrum (middle column) compared to a uniform spectrum, although they all relax to a final state around the same temperature anisotropy. This shows that high levels of a power spectrum may play a role on the regulation of the temperature anisotropies observed in different plasma environments. 

For a smaller value of $W_B(0)=0.03$ (orange lines in Fig.~\ref{fig:quasilinear}), the anisotropy also grows until it reaches a stationary state around $T_\perp/T_\parallel\simeq3$. However, this growth is monotonous and does not show a sharp increase nor a saturation in the early stages of the simulation, compared with $W_B(0)=0.1$. Similarly, $\beta_\parallel$ decreases almost monotonically from $0.1$ towards $0.05$. For even smaller values of the magnetic field intensity, e.g. $W_B(0)=0.01$, $0.003$, $0.001$ (green, red, and purple lines in Fig.~\ref{fig:quasilinear}), the anisotropy growth is limited and a stationary stage is reached at lower values near $T_\perp/T_\parallel\simeq1$. As $W_B(0)$ is lowered to noise levels $W_B(0)<10^{-5}$ (not shown), the anisotropy and other parameters remain almost constant, which is consistent with the fact that the plasma is in an equilibrium state for $\beta_\parallel=0.1$, $T_\perp/T_\parallel=1$, and low levels of the magnetic energy. In all cases, we observe that the total magnetic energy decreases monotonously, meaning that the quasilinear approximation is valid through every step of simulation runs. 

In the earlier stages of the simulation runs, most of the energy transfer from the waves to the particles occurs at $v_A{k_\parallel}/\Omega_p\gtrsim2$ since, according to the linear dispersion relation, the ACWs are heavily damped. This explains why a sufficient level of magnetic energy can heat the particles such that the anisotropy rises. Also, in this wavenumber range an initially uniform wave spectrum stores more energy compared to the Lorentzian one, meaning that the former can transfer more energy to particles compared to the latter in the same time lapse. A similar description holds as the Lorentzian is more energetic than the Gaussian in the $v_A{k_\parallel}/\Omega_p\gtrsim2$ range, explaining why the anisotropy can reach higher values for the uniform spectrum compared to the Lorentzian and Gaussian cases. After the field energy is exhausted for long wavenumbers, the anisotropy saturates. If this occurs for anisotropy values in which a kinetic instability is excited, which should happen around $v_A{k_\parallel}/\Omega_p=1$, then the energy transfer is reversed from the particles to the wave so that the anisotropy starts to decrease, and the wave energy around $v_A{k_\parallel}/\Omega_p=1$ grows at an instantaneous growth-rate $2\gamma$. However, as Fig.~\ref{fig:quasilinear} shows, this localized wave energy growth does not translate to total growth in $W_B$, probably because other parts of the spectrum are still transferring energy to the particles. If the saturated anisotropy is not enough to excite kinetic instabilities, or if the instability is weak, then energy transfer is a slow process and the plasma reaches a quasi-stationary state just after saturation, as shown in all cases with $W_B\leq0.03$ in Fig.~\ref{fig:quasilinear}.

\begin{figure}
  \centering
  \includegraphics[width=\linewidth]{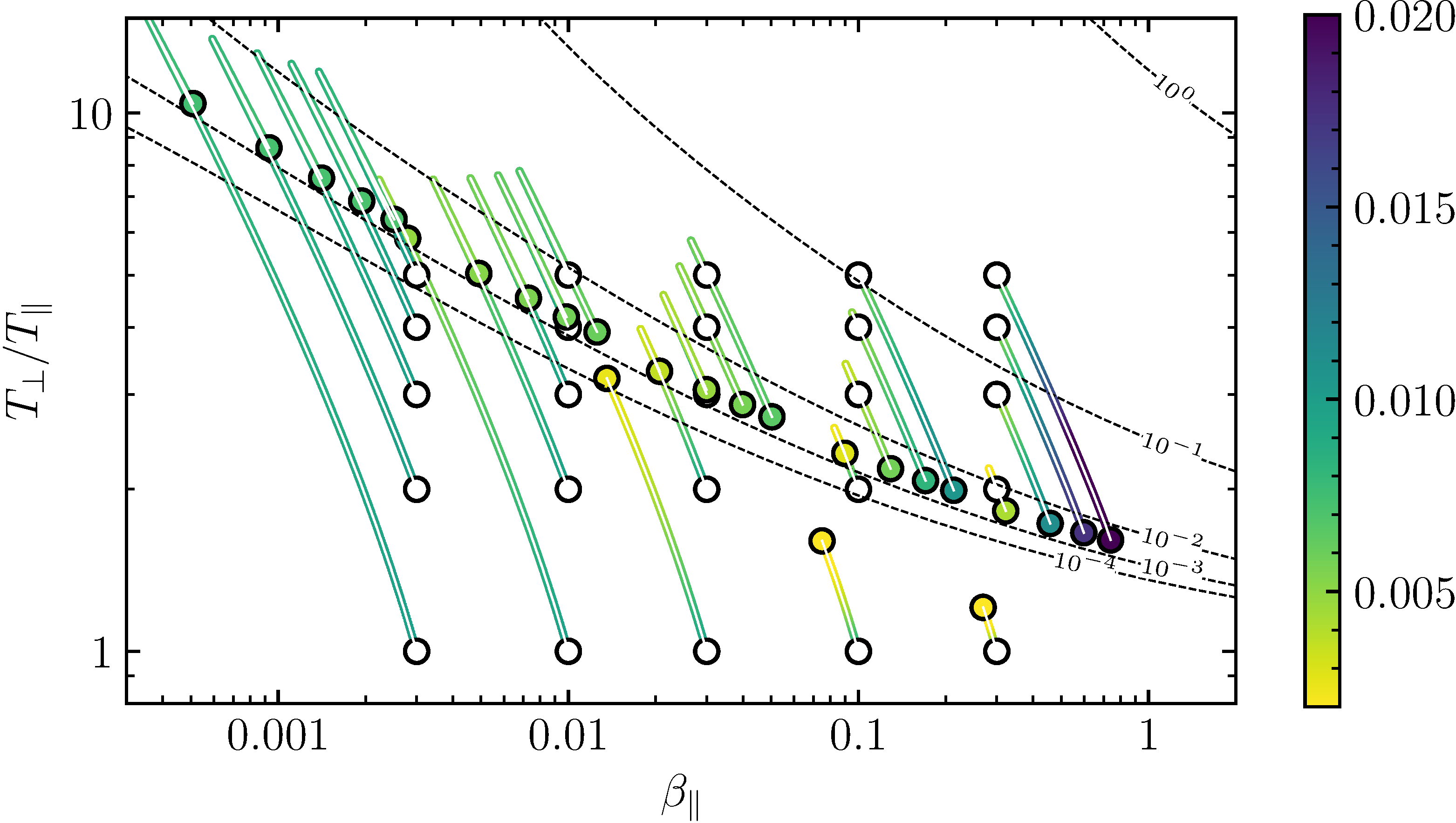}
  \caption{Quasilinear evolution of the proton $\beta_\parallel$ and anisotropy $T_\perp/T_\parallel$. Initial conditions (white circles) were chosen evenly spaced in the range $0.003\leq\beta_\parallel\leq0.3$ and $1\leq{T_\perp/T_\parallel}\leq5$, with a uniform magnetic wave spectrum of power $W_B(0)=0.01$ for all cases. The colorbar represents the instantaneous value of $W_B(t)$. Dashed lines are contours of the proton-cyclotron instability with maximum growth-rates  $\gamma/\Omega_p=10^{-4}$ through $10^0$, as calculated from the dispersion relation Eq.~\eqref{eq:disprel}. Colorized circles correspond to the final (stationary) state of the plasma simulations.}
  \label{fig:betaaniso}
\end{figure}

Figure~\ref{fig:betaaniso} shows the time evolution of $\beta_\parallel$, $T_\perp/T_\parallel$, and $W_B$. A set of numerical simulations with evenly spaced (in log space) initial conditions were chosen in the range $0.003\leq\beta_\parallel\leq0.3$ and $1\leq{T_\perp/T_\parallel}\leq5$, marked with white circles, with an initial uniform magnetic wave spectrum of power $W_B(0)=0.01$ for all cases. The colored lines represent the quasilinear path of the system in the diagram, and their colors represent the instantaneous magnetic energy $W_B$. The colored circles represent the final state of the system at $\Omega_p t=150$, long after the system has reached a stationary state, with color matching the instantaneous $W_B$.

A subset of these initial conditions can excite an instability as shown in Figs.~\ref{fig:disprel}(c) and (d). Thus, in Fig.~\ref{fig:betaaniso}, contours of the maximum growth-rate $\gamma_{\rm{}max}/\Omega_p$ are included as segmented lines. For numerical reasons, we will refer to the contour $\gamma/\Omega_p=10^{-4}$ as the instability threshold, and points below this line will be considered as stable according to the numerical solutions of the dispersion relation Eq.~\eqref{eq:disprel}. In general, the simulations with initial conditions below the instability threshold evolve so that the wave energy $W_B$ decreases monotonously from $W_B(0)=0.01$ to values between $0.002<W_B(t_{\rm{}sat})<0.01$, which is consistent with results from Fig.~\ref{fig:quasilinear}. On the other hand, for initially unstable conditions, the magnetic energy $W_B$ rises up to values below $0.01<W_B<0.02$.

In almost all simulated cases starting below the instability threshold, $\beta_\parallel$ drops rapidly while the temperature anisotropy increases to high values above the stability thresholds. Afterwards, the magnetic wave power is not enough to supply energy to protons, so that $T_\perp/T_\parallel$ slowly relaxes towards values where the maximum growth-rate of the ACW instability is of the order of $\gamma/\Omega_p\simeq0.001$. Of all the cases, however, there are three exceptions for $T_\perp(0)/T_\parallel(0)=1$ and $\beta_\parallel(0)=0.03$, $0.1$, and $0.3$, whose evolution reaches the final stationary state far below the instability threshold because the initial supplied energy $W_B(0)=0.1$ is insufficient to push the system to higher anisotropies. This may explain why most of the observed plasma parameters in the solar wind are close to $\beta_\parallel=1$ and $T_\perp/T_\parallel=1$, since plasmas in this state are not heavily influenced by the background spectrum.

Notice that for simulations starting far below or far above the instability threshold, e.g. $\beta_\parallel(0)=0.003$ and $T_\perp/T_\parallel(0)=1$, or $\beta_\parallel(0)=0.3$ and $T_\perp/T_\parallel(0)=5$, the anisotropy can grow up to very high values above the instability thresholds. This effect is damped as the starting anisotropy is near the instability threshold. Thus, the effects of the starting anisotropy, which is a measure of the free energy available to excite an  instability, can compete with the effects of the starting magnetic energy to regulate the anisotropy growth. Although not shown here, the quasilinear evolution in the cases of a Gaussian or Lorentzian power spectrum are similar. They all excite some level of proton perpendicular heating in the initial stage of the simulations, and then relax slowly towards a quasi-stationary state near the instability threshold, with properties similar to the ones shown in Fig.~\ref{fig:quasilinear}.

In summary, we have illustrated how the initial shape magnetic field background spectrum can produce different results on the evolution of macroscopic parameters of the plasma. However, solar wind observations show that the plasma is mostly in a state below the instability thresholds, far from the isotropic state~\cite{kasper2002,Hellinger2006}, with a non-negligible level of magnetic fluctuations~\cite{bale2009,Navarro2014}, and that the magnetic field has a spectral break around the ion inertial length~\citep{Chen2014,Wang2018}. The inertial range for transverse fluctuations propagating along the magnetic field $v_A k_\parallel/\Omega_p<1$ typically shows a power-law spectrum $B_k^2\propto k_\parallel^{-2}$~\cite{Horbury2008}. For ion or sub-ions scales (in the kinetic range) the turbulent spectrum steepens to $k_\parallel^{-\alpha}$, with $\alpha\ge2.0$, arguably due to the characteristics of the dispersion relation of Alfv\'en or other waves in that range~\citep{Galtier2003,Gary2009,Schekochihin2009,Boldyrev2013}. Thus, the results presented here for three arbitrary spectral shapes may not remain the same when a solar wind-like spectrum is considered. This will be the focus of the next section.

\section{Numerical Results. The effect of a turbulent spectrum with a spectral break}
\label{sec:Results2}

Here we compute the quasilinear relaxation considering a quasi-parallel solar wind-like spectrum, including a spectral break at the ion inertial range scale, given by:
\begin{equation}
\label{eq:alpha}
    |B_k(0)|^2 = 
    \begin{cases}
    A k_\parallel^{-2} & v_A k_\parallel/\Omega_p < 1 \,, \\
    A k_\parallel^{-\alpha} & v_A k_\parallel/\Omega_p > 1 \,,
    \end{cases}
\end{equation}
where $A$ is chosen depending on the initial total magnetic energy $W_B(0)$, with the integral calculated in the range $10^{-3}< v_A |k_\parallel|/\Omega_p < 8$. Notice that we are restricting to a reduced 1D background spectrum in $k_{\parallel}$. In this case, a $k_{\parallel}^{-2}$ reduced spectrum in the MHD range is the result of the integration in $k_{\perp}$ of a 2D spectrum of quasi-perpendicular fluctuations, whose corresponding reduced $k_{\perp}$ spectrum exhibit a $-5/3$ power law and a spectral anisotropy of $k_{\parallel} \sim k_{\perp}^{2/3}$. In general, the spectral slopes are determined by the conservation of total energy, i.e. that $\int dk_{\parallel}\,E(k_{\parallel}) =\int dk_{\perp}\,E(k_{\perp})$; and by the assumed spectral anisotropy, e.g. $k_{\parallel} \sim k_{\perp}^{d/3}$. So, if the reduced perpendicular 1D spectrum is $k_{\perp}^{-\mu}$, then the corresponding parallel reduced spectrum scales as $k_{\parallel}^{-\alpha}$ with $\alpha = [3(\mu-1) + d]/d$. The spectral anisotropy (represented by the parameter $d$) is still a matter of great debate. Typical values of such anisotropy are $d = 1$ (standard kinetic-Alfvén-wave (KAW) turbulence  for which $\mu=7/3$, giving $\alpha=5$), $d=2$ (intermittency corrected KAW turbulence~\citep{Boldyrev2012}, for which $\mu=8/3$, giving $\alpha=7/2$), or some reconnection-mediated scenario where $d=3$ (i.e. $k_{\parallel} \sim k_{\perp}$, to which $\alpha=\mu$). Sometimes it has been found $\alpha=\mu=3$ in simulations~\citep{Arzamasskiy2019}.

Therefore, following the several observations mentioned here and in the Introduction section, we have considered four values of the spectral index, namely $\alpha=2,\,7/3,\,7/2$, and $5$. Notice that the case with $\alpha=2$ corresponds to fluctuations without a break spectrum, which is unrealistic as a break is always observed around the ion characteristic scales and the spectrum is always steeper at smaller scales. Nevertheless we include such case for comparison purposes. Also notice that existing theories of plasma turbulence predict power laws in $k_\parallel$ with e.g. $\alpha=2$ in the MHD range (see e.g. \citet{Howes2015}), in consistency with solar wind observations~\cite{Horbury2008}, although this heavily relies on the assumed spectral anisotropy of the turbulent fluctuations, which is still a matter of great debate when it comes to the kinetic range. 
Moreover, regarding the validity of a purely parallel (instead of quasi-parallel) model, it is important to mention that, as shown by ~\citet{Gaelzer2015,Kim2016}, results considering quasi-parallel propagation may differ only by a multiplicative scaling factor with respect to the purely parallel propagation case considered in this work.
 
\begin{figure}
  \centering
  \includegraphics[width=\textwidth]{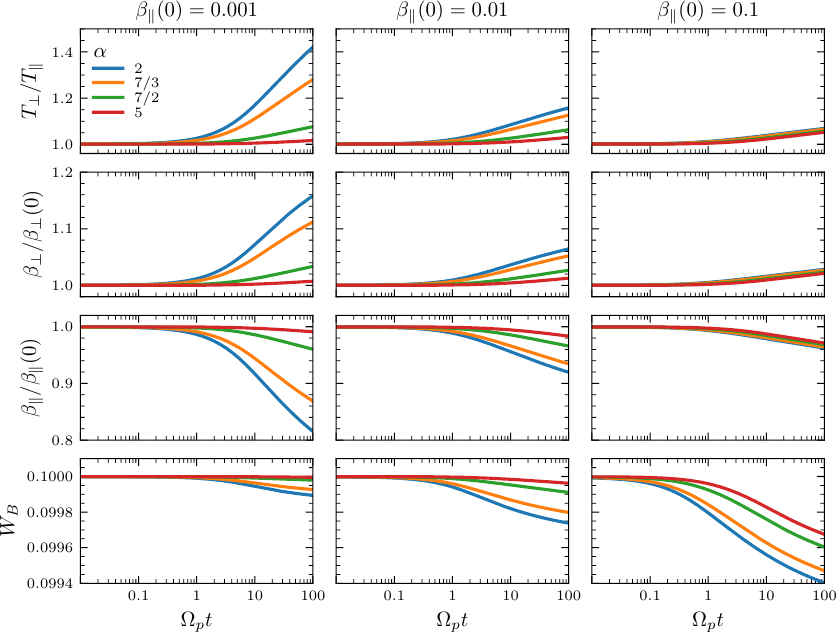}
  \caption{Quasilinear evolution of $T_\perp/T_\parallel$, $\beta_\perp$, $\beta_\parallel$, and total magnetic energy $W_B$ for a solar-wind like spectrum Eq.~\eqref{eq:alpha} with different values of $2\leq\alpha\leq5$. Initially, $T_\perp/T_\parallel(0)=1$, $W_B(0)=0.1$, and (left column) $\beta_\parallel(0)=0.001$, (middle column) $\beta_\parallel(0)=0.01$, and (right column) $\beta_\parallel(0)=0.1$. }
  \label{fig:alpha}
\end{figure}

In what follows, the initial anisotropy and total magnetic energy are chosen as $T_\perp(0)/T_\parallel(0)=1$ and $W_B(0)=0.1$ for all simulation runs. For an initially low $\beta_\parallel(0)=0.001$, Fig.~\ref{fig:alpha} (left column) shows that the proton distribution is cooled in the parallel direction with respect to the background magnetic field, as the parallel $\beta_\parallel$ decreases in time. Similarly, protons are heated in the transverse direction for all tested values of $\alpha$. It is interesting to note that the magnetic energy decreases just 1\% from the initial value, but causing a monotonous growth in the temperature anisotropy from $T_\perp/T_\parallel=1$ to $\simeq1.4$ for $\alpha=2$. For larger values of $\alpha$, this parallel cooling and transverse heating is less efficient. This can be explained as a steepened magnetic field spectrum for $v_Ak_\parallel/\Omega_p>1$ do not contain enough energy to be transferred to the particles compared to the $\alpha=2$ case. 

\begin{figure}
  \centering
  \includegraphics[width=\textwidth]{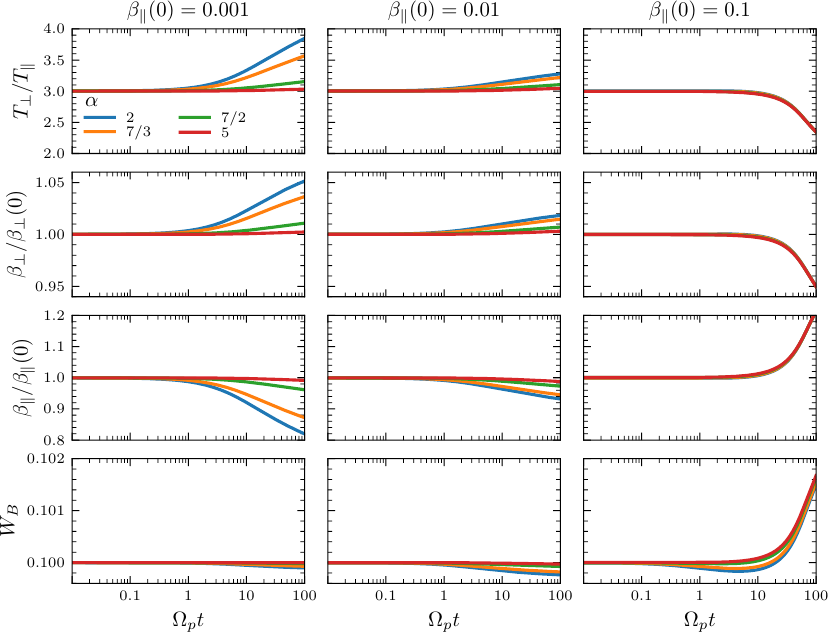}
  \caption{Same as in Fig.~\ref{fig:alpha} but for an initial anisotropy $T_\perp/T_\parallel(0)=3$. }
  \label{fig:alpha_R3}
\end{figure}

For $\beta_\parallel(0)=0.01$ and $0.1$, we see in Fig.~\ref{fig:alpha} that the parallel cooling and transverse heating still occurs. It is worth noticing that the magnetic energy actually decreases more, but the parallel cooling and transverse heating is less efficient than in cases with the same value of $\alpha$ and lower $\beta_\parallel(0)=0.001$. Comparing with Fig.~\ref{fig:betaaniso}, we see that all three cases $\beta_\parallel=0.001$, $0.01$, and $0.1$, with $T_\perp/T_\parallel=1$, correspond to linearly stable plasmas. However, $\beta_\parallel=0.1$ is closer to the instability thresholds, meaning that the quasilinear evolution will likely reach a stationary state with lower anisotropies near the stability margins. Also, in the cases we tested for high $\beta_\parallel(0)\geq0.1$, the temperature evolution seems to be independent of $\alpha$. Moreover, as an initially anisotropic plasma can drive instabilities depending on the plasma beta, in order to compare with the initially isotropic case (always stable) shown in Fig.~\ref{fig:alpha}, Fig.~\ref{fig:alpha_R3} shows results for the same set of parameters as in Fig.~\ref{fig:alpha}, but for an initially anisotropic plasma with $T_\perp/T_\parallel(0)=3$. The cases with $\beta_\parallel(0)=0.001$ and $0.01$ show similar qualitative characteristics for both Figs.~\ref{fig:alpha} and \ref{fig:alpha_R3}. However, the case with $\beta_\parallel(0)=0.1$ is initially unstable [see the green line in Fig.~\ref{fig:disprel}(c)]. This results in the reduction of the anisotropy and the increase in $\beta_\parallel$.      

\begin{figure}
  \centering
  \includegraphics[width=\textwidth]{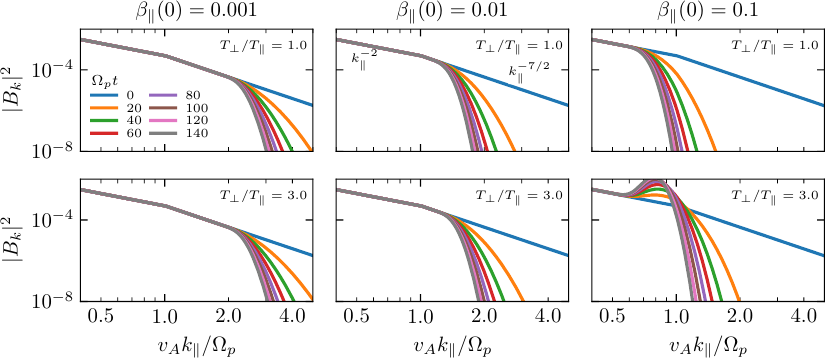}
  \caption{Spectral wave energy $|B_k|^2$ as a function of the normalized wavenumber $v_Ak_\parallel/\Omega_p$ for several time steps. Initial conditions are $W_B(0)=0.1$ for a solar wind-like spectrum with $\alpha=7/2$. (top) $T_\perp/T_\parallel(0)=1$ and (bottom) $T_\perp/T_\parallel(0)=3$. (left) $\beta_\parallel(0)=0.001$, (middle) $\beta_\parallel(0)=0.01$, and (right) $\beta_\parallel(0)=0.1$. }
  \label{fig:bk}
\end{figure}
Finally, Fig.~\ref{fig:bk} shows the spectral wave energy density at several intermediate time steps between $\Omega_p t=0$ and $\Omega_pt=140$, considering the initial conditions $W_B(0)=0.1$, $\alpha=7/2$, $T_\perp(0)/T_\parallel(0)=1$ (top),  $T_\perp(0)/T_\parallel(0)=3$ (bottom), and the same three values of $\beta_\parallel(0)$ as in Figs.~\ref{fig:alpha} and~\ref{fig:alpha_R3}. For other values of $\alpha$ the description of these figures are almost identical. As time goes on, the wave spectrum is dampened for high values of $k_\parallel$ in all cases, which is consistent with results of Fig.~\ref{fig:disprel}. For $\beta_\parallel(0)=0.001$, the spectral break is unmodified at all times of the simulation run, but the spectrum steepens for $v_Ak_\parallel/\Omega_p>2$ for both $T_\perp(0)/T_\parallel(0)=1$ and $3$. For $\beta_\parallel(0)=0.01$ and $T_\perp(0)/T_\parallel(0)=1$ and $3$, this happens at $v_Ak_\parallel/\Omega_p>1.2$ and the spectral break is still visible. For $\beta_\parallel(0)=0.1$, the spectral break disappears and the wave spectrum becomes smooth around $v_Ak_\parallel/\Omega_p=1$. In all these cases, transference of energy from the wave to protons results in a monotonous drop in magnetic energy as discussed in Fig.~\ref{fig:alpha} and~\ref{fig:alpha_R3}. 

For $\beta_\parallel(0)=0.1$ and $T_\perp/T_\parallel(0)=3$, in the first stages of the simulation the electromagnetic wave loses energy at high values of $v_Ak_\parallel/\Omega_p>1$ as in the previous cases. However, since the wave is unstable in this case around $0.6<v_Ak_\parallel/\Omega_p\lesssim1.2$ [see green line in Fig.~\ref{fig:disprel}(c)], then the magnetic field amplitude starts to grow for those wavenumbers resulting in a bump in the spectral wave energy just below the spectral break. This results in the decrease of the temperature anisotropy and increment of $\beta_\parallel$, as shown in Fig.~\ref{fig:alpha_R3}, which is also consistent with the discussion of Fig.~\ref{fig:betaaniso}. Comparing with Fig.~\ref{fig:disprel}, this implies that the range of unstable wavenumbers shifts towards smaller values, which in turns means that the bump in the spectral wave energy also shifts to smaller values of $k_\parallel$. At the same time, previously unstable modes with higher $v_Ak_\parallel/\Omega_p>1$ become damped. Thus the wave transfers energy to protons for values of $v_A k_\parallel/\Omega_p>1$, resulting in a steep spectrum near the initial spectral break. At larger times and since the rate at which the wave damps is negligible compared to its growth, this results in a total growth of magnetic energy as shown in the lower right panel of Fig.~\ref{fig:alpha_R3}, which is consistent with the description of Fig.~\ref{fig:betaaniso} for initially unstable plasma states.

\section{Discussion and conclusions}
\label{sec:conclusions}

Turbulence is ubiquitous in space environments and any relaxation process should occur in the presence of a background turbulent magnetic spectrum, e.g. relaxation to quasi-stationary states out of thermodynamic equilibrium, non-Maxwellian characteristics in poorly collisional plasmas, and temperature anisotropy regulation by micro-instabilities or other processes, among others. Here, we have focused on the possible effects of a magnetic field background spectrum on the quasilinear relaxation of the Alfv\'en ion-cyclotron temperature anisotropy instability. We have compared four different choices of the initial magnetic field spectrum shape $|B_k|^2$: (a) uniform noise, (b) Lorentzian, (c) Gaussian, and (d) the more realistic solar wind-like spectrum with a spectral break at $v_Ak_\parallel/\Omega_p=1$ and the kinetic ion or sub-ion range $v_Ak_\parallel/\Omega_p>1$ with a spectrum $\propto k^{-\alpha}$ with varying $\alpha \ge 2$.

Considering a plasma composed of bi-Maxwellian protons and cold electrons, with electromagnetic fluctuations propagating along a background magnetic field $\vec{B}_0$, it is shown that all the spectrum shapes considered here can heat protons preferentially in the direction perpendicular with respect to $\vec{B}_0$, provided the initial wave has enough energy power to be transferred to protons, even if the plasma is initially stable. Thus, isotropic protons can be heated towards high anisotropies $T_\perp/T_\parallel > 1$. If $T_\perp/T_\parallel$ reaches high enough values, then an Alfv\'en-cyclotron instability is excited during the quasilinear evolution. Afterwards, the anisotropy production saturates and the plasma relaxes to a quasi-stationary state with a maximum instability growth-rate $\gamma/\Omega_p\approx10^{-3}$. If the initial wave energy is insufficient then kinetic instabilities are not excited, although transverse heating may still occur. 

It is noted that for simulations starting far below the instability threshold (defined here as the contour where $\gamma_{\rm{}max}/\Omega_p=10^{-4}$ in the $\beta_\parallel$ and $T_\perp/T_\parallel$ parameter space), the anisotropy grows, and  $\beta_\parallel$ and the wave energy are reduced, such that the final quasi-stationary state lies near the instability thresholds. This means that simulations of stable protons ($T_\perp/T_\parallel=1$) starting with low $\beta_\parallel<0.01$ experience stronger perpendicular heating compared to simulations starting with $\beta_\parallel>0.01$ for the same initial wave energy $W_B$. For higher values of $\beta_\parallel>0.1$, a wave energy $W_B<0.1$ may not be sufficient to be transferred to protons, as it must compete with kinetic effects as measured by the beta parameter, thus the quasi-stationary state is reached for anisotropy values far below the instability thresholds although transverse heating can still be observed. On the other hand, simulations where the plasma is initially unstable (for anisotropies above the instability thresholds), the kinetic instability is dominant over the effects of energy transfer from the turbulent wave to protons. This results in an effective transverse cooling, i.e. reduction of the anisotropy and growth of $\beta_\parallel$. The wave energy also grows due to the presence of the instability. Nevertheless, and similarly to the previous description, the quasilinear evolution reaches a stationary state around the instability thresholds. Thus, there exists competition between the energy transfer from turbulent waves and the presence of kinetic instabilities, and they are effectively balanced near the instability thresholds.

Our numerical results show that the proton transverse heating by the waves is efficient depending on the energy stored in the tail of the magnetic spectrum ($v_Ak_\parallel/\Omega_p>1$ which lies in the kinetic range). This happens because the waves are heavily damped in the kinetic range according to the linear kinetic dispersion relation, and energy transfer from the waves to protons should occur first for those wavenumbers according to the quasilinear description. As anticipated from the previous results, in the case of a solar wind-like turbulent power-law spectra with a spectral break, the perpendicular heating is more effective for lower values of $\alpha$ as the tails of the spectrum can store more energy to be transferred to the particles. For values of $\beta_\parallel(0)>0.01$, transverse preferential heating still occurs, however it seems to be independent of $\alpha$, and it is less efficient than for lower values of $\beta_\parallel(0)<0.01$. This means that for high values of $\beta_\parallel$, the shape of the wave spectrum is less relevant for perpendicular heating, and other kinetic effects become dominant. Finally, if the plasma is initially unstable, i.e. with large enough temperature anisotropy (e.g. $T_\perp/T_\parallel=3$ for $\beta_\parallel=0.1$), then the wave spectrum grows in time because it absorbs energy from the particles in a range of wavenumbers $v_Ak_\parallel/\Omega_p<1$. This produces a bump in the spectral wave energy just below the spectral break, the growth of the total magnetic energy, and the subsequent reduction of the proton temperature anisotropy towards the instability thresholds. It is worth mentioning that a bump has been sometimes observed in the magnetic field spectrum in correspondence of the ion characteristic scales near the Sun~\citep{Bowen2020}. Thus, our results suggest that such characteristic may be related with resonant wave-particle interactions between unstable ion populations and turbulence near the spectral break.

The above being said, it is important to mention that turbulence in the solar wind correspond to an anisotropic cascade following critical balance with $k_\perp \neq 0$. However, as a first approximation here we have considered the fluctuations to follow the parallel propagating ($k_\perp=0$) Alfv\'en Ion-Cyclotron mode dispersion relation, since for small values of plasma beta, the compressibility of the fluctuations is small as pointed out by~\citet{bale2009}, which is consistent with Alfv\'en Ion-Cyclotron modes. Under this context we have considered that the plasma will only interact with transverse fluctuations with respect to the mean field, and therefore we have neglected the effect of other fluctuations with $k_\perp \neq 0$. We recognize, however, that this may be considered a crude approximation but at the same time we believe that our simplified approach provides valuable insights to the problem of the heating of the solar wind, that is generally observed in an anisotropic state. We expect to expand the scope of our approach and results with a subsequent study considering oblique propagating waves, hopefully corroborating or improving the results of our reduced model.

In summary, our results suggest a possible mechanism to explain why the solar wind plasma can be observed in a stationary state with $T_\perp/T_\parallel >1$ near the instability thresholds or far from thermodynamic equilibrium, as has been observed in the Earth's magnetosphere or the solar wind. A sufficient level of turbulent magnetic spectrum can drive an initially stable proton plasma towards higher values of the temperature anisotropy, i.e. far from thermodynamic equilibrium and preventing the plasma to remain in an isotropic state.  However, measurements of solar wind protons at different space environments show that proton velocity distributions can also exhibit anisotropic states with $T_\perp/T_\parallel <1$~\citep{bale2009}. There are several possible explanations for this apparent discrepancy, as in this work we have not considered other effects that can effectively reduce the production of anisotropy, or compete with the Alfv\'en-cyclotron instability and the turbulent energy transfer, but are nonetheless worth of study. For example, radial expansion from the Sun~\citep{Moya2012,Hellinger2015,Yoon2017b}, collisional  effects~\citep{Yoon2017}, the role of compressive fluctuations~\citep{yan2011}, oblique propagation and the corresponding anisotropic turbulent cascade~\citep{Horbury2005,Horbury2008,Howes2015}, oblique instabilities such as the mirror or oblique firehose instabilities~\citep{yoon}, other non-Maxwellian distributions such as kappa distributions, the presence of heavy ions~\citep{Navarro2020}, or the kinetic effects of electrons. Also, the amplitude of magnetic fluctuations decays mostly with $\beta_\parallel$~\citep{bale2009,Navarro2014}, imposing a severe restriction on the wave energy available for anisotropy production in space plasmas. However, steep spectra (larger $\alpha$ values) have been typically observed in association with small values of the plasma beta and large levels of turbulent fluctuations, both recently in near-Sun environment by Parker Solar Probe, and also previously in the near-Earth environment by WIND~\citep{Bruno2014}. Moreover, Parker Solar Probe measurements have also shown an enhanced perpendicular proton heating possibly due to stochastic heating related to the strong turbulent fluctuations particularly in the fast solar wind (see e.g.~\citet{Martinovic2020}) that could compete with all the mentioned mechanisms, including the heating mechanism suggested in this paper.

\section*{Conflict of Interest Statement}

The authors declare that the research was conducted in the absence of any commercial or financial relationships that could be construed as a potential conflict of interest.

\section*{Author Contributions}

Both authors planned, outlined, researched, and wrote the manuscript. 

\section*{Funding}
This work was supported by ANID, Chile, through FONDECyT grants No. 11180947 (R.E.N.), and No. 1191351 (P.S.M), and CONICyT-PAI grant No. 79170095 (R.E.N).


\section*{Data Availability Statement}
The original contributions presented in the study are included in the article/supplementary material, further inquiries can be directed to the corresponding author/s.

\bibliographystyle{frontiersinHLTH&FPHY} 
\bibliography{references}

\begin{thebibliography}{83}
\expandafter\ifx\csname natexlab\endcsname\relax\def\natexlab#1{#1}\fi
\expandafter\ifx\csname urlstyle\endcsname\relax
  \expandafter\ifx\csname doi\endcsname\relax
  \def\doi#1{doi:\discretionary{}{}{}#1}\fi \else
  \expandafter\ifx\csname doi\endcsname\relax
  \def\doi{doi:\discretionary{}{}{}\begingroup \urlstyle{rm}\Url}\fi \fi
\expandafter\ifx\csname selectlanguage\endcsname\relax
  \def\selectlanguage#1{}\fi

\bibitem[{Spitzer(1962)}]{spitzer1962}
Spitzer L.
\newblock {\em Physics of fully ionized gases\/} (New York: John Wiley \& Sons,
  Inc.) (1962).

\bibitem[{Marsch and Goldstein(1983)}]{marsch1983}
Marsch E, Goldstein H.
\newblock The effects of {Coulomb} collisions on solar wind ion velocity
  distributions.
\newblock {\em J. Geophys. Res.\/} {\bf 88} (1983) 9933--9940.
\newblock \doi{10.1029/JA088iA12p09933}.

\bibitem[{Marsch et~al.(1982)Marsch, M{\"u}hlh{\"a}user, Rosenbauer, Schwenn,
  and Neubauer}]{Marsch1982}
Marsch E, M{\"u}hlh{\"a}user KH, Rosenbauer H, Schwenn R, Neubauer F.
\newblock Solar wind helium ions: Observations of the helios solar probes
  between 0.3 and 1 au.
\newblock {\em J. Geophys. Res.\/} {\bf 87} (1982) 35--51.
\newblock \doi{10.1029/JA087iA01p00035}.

\bibitem[{Gary(1993)}]{gary1993}
Gary SP.
\newblock {\em Theory of Space Plasma Microinstabilities\/} (Cambridge
  University Press) (1993).
\newblock \doi{10.1017/CBO9780511551512}.

\bibitem[{Astudillo(1996)}]{Astudillo1996}
Astudillo HF.
\newblock High-order modes of left-handed electromagnetic waves in a
  solar-wind-like plasma.
\newblock {\em J. Geophys.\ Res.\/} {\bf 101} (1996) 24433.

\bibitem[{Hellinger and Tr{\'{a}}vn{\'{i}}{\v{c}}ek(2006)}]{Hellinger2006}
Hellinger P, Tr{\'{a}}vn{\'{i}}{\v{c}}ek P.
\newblock {Parallel and oblique proton fire hose instabilities in the presence
  of alpha/proton drift: Hybrid simulations}.
\newblock {\em J. Geophys. Res.\/} {\bf 111} (2006) A01107.
\newblock \doi{10.1029/2005JA011318}.

\bibitem[{L{\'o}pez et~al.(2020)L{\'o}pez, Lazar, Shaaban, Poedts, and
  Moya}]{Lopez2020}
L{\'o}pez RA, Lazar M, Shaaban SM, Poedts S, Moya PS.
\newblock Alternative high-plasma beta regimes of electron heat-flux
  instabilities in the solar wind.
\newblock {\em Astrophys. J. Lett.\/} {\bf 900} (2020) L25.
\newblock \doi{10.3847/2041-8213/abaf56}.

\bibitem[{Klimontovich(1997)}]{klimontovich1997}
Klimontovich YL.
\newblock Physics of collisionless plasma.
\newblock {\em Physics-Uspekhi\/} {\bf 40} (1997) 21.

\bibitem[{Kasper et~al.(2002)Kasper, Lazarus, and Gary}]{kasper2002}
Kasper JC, Lazarus AJ, Gary SP.
\newblock {Wind/SWE observations of firehose constraint on solar wind proton
  temperature anisotropy}.
\newblock {\em Geophys. Res. Lett.\/} {\bf 29} (2002) 20--1--20--4.
\newblock \doi{10.1029/2002GL015128}.

\bibitem[{Marsch(2006)}]{marsch2006}
Marsch E.
\newblock Kinetic physics of the solar corona and solar wind.
\newblock {\em Living Rev. Solar Phys.\/} {\bf 3} (2006).
\newblock \doi{10.12942/lrsp-2006-1}.

\bibitem[{Bale et~al.(2009)Bale, Kasper, Howes, Quataert, Salem, and
  Sundkvist}]{bale2009}
Bale SD, Kasper JC, Howes GG, Quataert E, Salem C, Sundkvist D.
\newblock Magnetic fluctuation power near proton temperature anisotropy
  instability thresholds in the solar wind.
\newblock {\em Phys.\ Rev.\ Lett.\/} {\bf 103} (2009) 211101.
\newblock \doi{10.1103/PhysRevLett.103.211101}.

\bibitem[{Bruno and Carbone(2013)}]{bruno2013}
Bruno R, Carbone V.
\newblock The solar wind as a turbulence laboratory.
\newblock {\em Living Rev. Solar Phys.\/} {\bf 10} (2013) 2.
\newblock \doi{10.12942/lrsp-2013-2}.

\bibitem[{Heinemann(1999)}]{heinemann1999}
Heinemann M.
\newblock Role of collisionless heat flux in magnetospheric convection.
\newblock {\em Journal of Geophysical Research: Space Physics\/} {\bf 104}
  (1999) 28397--28410.
\newblock \doi{10.1029/1999JA900401}.

\bibitem[{Matsumoto and Hoshino(2006)}]{matsumoto2006}
Matsumoto Y, Hoshino M.
\newblock Turbulent mixing and transport of collisionless plasmas across a
  stratified velocity shear layer.
\newblock {\em Journal of Geophysical Research: Space Physics\/} {\bf 111}
  (2006).
\newblock \doi{10.1029/2004JA010988}.

\bibitem[{Espinoza et~al.(2018)Espinoza, Stepanova, Moya, Antonova, and
  Valdivia}]{espinoza2018}
Espinoza CM, Stepanova M, Moya PS, Antonova EE, Valdivia JA.
\newblock Ion and electron $\kappa$ distribution functions along the plasma
  sheet.
\newblock {\em Geophys. Res. Lett.\/} {\bf 45} (2018) 6362--6370.
\newblock \doi{10.1029/2018GL078631}.

\bibitem[{Moya et~al.(2015)Moya, Pinto, Vi{\~n}as, Sibeck, Kurth, Hospodarsky
  et~al.}]{moya2015}
Moya PS, Pinto VA, Vi{\~n}as AF, Sibeck DG, Kurth WS, Hospodarsky GB, et~al.
\newblock {Weak kinetic Alfv{\'e}n waves turbulence during the
  14 November 2012 geomagnetic storm: Van Allen Probes observations}.
\newblock {\em Journal of Geophysical Research: Space Physics\/} {\bf 120}
  (2015) 5504--5523.
\newblock \doi{10.1002/2014JA020281}.

\bibitem[{Narita et~al.(2020)Narita, Roberts, V{\"o}r{\"o}s, and
  Hoshino}]{Narita2020}
Narita Y, Roberts OW, V{\"o}r{\"o}s Z, Hoshino M.
\newblock Transport ratios of the kinetic alfv{\'e}n mode in space plasmas.
\newblock {\em Frontiers in Physics\/} {\bf 8} (2020) 166.
\newblock \doi{10.3389/fphy.2020.00166}.

\bibitem[{Sahraoui et~al.(2010)Sahraoui, Goldstein, Belmont, Canu, and
  Rezeau}]{Sahraoui2010}
Sahraoui F, Goldstein ML, Belmont G, Canu P, Rezeau L.
\newblock Three dimensional anisotropic {\$}k{\$} spectra of turbulence at
  subproton scales in the solar wind.
\newblock {\em Physical Review Letters\/} {\bf 105} (2010) 131101.
\newblock \doi{10.1103/PhysRevLett.105.131101}.

\bibitem[{Horbury et~al.(2008)Horbury, Forman, and Oughton}]{Horbury2008}
Horbury TS, Forman M, Oughton S.
\newblock Anisotropic scaling of magnetohydrodynamic turbulence.
\newblock {\em Phys. Rev. Lett.\/} {\bf 101} (2008) 175005.
\newblock \doi{10.1103/PhysRevLett.101.175005}.

\bibitem[{Chen(2016)}]{Chen2016}
Chen CHK.
\newblock Recent progress in astrophysical plasma turbulence from solar wind
  observations.
\newblock {\em Journal of Plasma Physics\/} {\bf 82} (2016) 535820602.
\newblock \doi{10.1017/S0022377816001124}.
\newblock 535820602.

\bibitem[{Gonz{\'a}lez et~al.(2019)Gonz{\'a}lez, Parashar, Gomez, Matthaeus,
  and Dmitruk}]{Gonzalez2019}
Gonz{\'a}lez CA, Parashar TN, Gomez D, Matthaeus WH, Dmitruk P.
\newblock Turbulent electromagnetic fields at sub-proton scales: Two-fluid and
  full-kinetic plasma simulations.
\newblock {\em Physics of Plasmas\/} {\bf 26} (2019) 012306.
\newblock \doi{10.1063/1.5054110}.

\bibitem[{Cerri et~al.(2019)Cerri, Gro{\v{s}}elj, and Franci}]{Cerri2019}
Cerri SS, Gro{\v{s}}elj D, Franci L.
\newblock Kinetic plasma turbulence: Recent insights and open questions from
  3d3v simulations.
\newblock {\em Frontiers in Astronomy and Space Sciences\/} {\bf 6} (2019) 64.
\newblock \doi{10.3389/fspas.2019.00064}.

\bibitem[{Howes et~al.(2008)Howes, Cowley, Dorland, Hammett, Quataert, and
  Schekochihin}]{Howes2008}
Howes GG, Cowley SC, Dorland W, Hammett GW, Quataert E, Schekochihin AA.
\newblock A model of turbulence in magnetized plasmas: Implications for the
  dissipation range in the solar wind.
\newblock {\em Journal of Geophysical Research: Space Physics\/} {\bf 113}
  (2008).
\newblock \doi{10.1029/2007JA012665}.

\bibitem[{Maneva et~al.(2015)Maneva, Vi{\~n}as, Moya, Wicks, and
  Poedts}]{maneva2015}
Maneva Y, Vi{\~n}as AF, Moya PS, Wicks RT, Poedts S.
\newblock Dissipation of parallel and oblique alfv{\'e}n-cyclotron
  waves--implications for heating of alpha particles in the solar wind.
\newblock {\em Astrophys. J.\/} {\bf 814} (2015) 33.

\bibitem[{Sulem and Passot(2015)}]{sulem2015}
Sulem PL, Passot T.
\newblock Landau fluid closures with nonlinear large-scale finite larmor radius
  corrections for collisionless plasmas.
\newblock {\em Journal of Plasma Physics\/} {\bf 81} (2015) 325810103.
\newblock \doi{10.1017/S0022377814000671}.

\bibitem[{Cerri and Califano(2017)}]{Cerri2017}
Cerri SS, Califano F.
\newblock Reconnection and small-scale fields in 2d-3v hybrid-kinetic driven
  turbulence simulations.
\newblock {\em New Journal of Physics\/} {\bf 19} (2017) 025007.
\newblock \doi{10.1088/1367-2630/aa5c4a}.

\bibitem[{Franci et~al.(2017)Franci, Cerri, Califano, Landi, Papini, Verdini
  et~al.}]{Franci2017}
Franci L, Cerri SS, Califano F, Landi S, Papini E, Verdini A, et~al.
\newblock Magnetic reconnection as a driver for a sub-ion-scale cascade in
  plasma turbulence.
\newblock {\em The Astrophysical Journal\/} {\bf 850} (2017) L16.
\newblock \doi{10.3847/2041-8213/aa93fb}.

\bibitem[{Loureiro and Boldyrev(2017)}]{Loureiro2017}
Loureiro NF, Boldyrev S.
\newblock Collisionless reconnection in magnetohydrodynamic and kinetic
  turbulence.
\newblock {\em The Astrophysical Journal\/} {\bf 850} (2017) 182.
\newblock \doi{10.3847/1538-4357/aa9754}.

\bibitem[{Mallet et~al.(2017)Mallet, Schekochihin, and Chandran}]{mallet2017}
Mallet A, Schekochihin AA, Chandran BDG.
\newblock Disruption of alfvénic turbulence by magnetic reconnection in a
  collisionless plasma.
\newblock {\em Journal of Plasma Physics\/} {\bf 83} (2017) 905830609.
\newblock \doi{10.1017/S0022377817000812}.

\bibitem[{Papini et~al.(2019)Papini, Franci, Landi, Verdini, Matteini, and
  Hellinger}]{Papini2019}
Papini E, Franci L, Landi S, Verdini A, Matteini L, Hellinger P.
\newblock Can hall magnetohydrodynamics explain plasma turbulence at sub-ion
  scales?
\newblock {\em The Astrophysical Journal\/} {\bf 870} (2019) 52.
\newblock \doi{10.3847/1538-4357/aaf003}.

\bibitem[{Chen et~al.(2014)Chen, Leung, Boldyrev, Maruca, and Bale}]{Chen2014}
Chen CHK, Leung L, Boldyrev S, Maruca BA, Bale SD.
\newblock Ion-scale spectral break of solar wind turbulence at high and low
  beta.
\newblock {\em Geophysical Research Letters\/} {\bf 41} (2014) 8081--8088.
\newblock \doi{10.1002/2014GL062009}.

\bibitem[{Wang et~al.(2018)Wang, Tu, He, and Wang}]{Wang2018}
Wang X, Tu CY, He JS, Wang LH.
\newblock Ion-scale spectral break in the normal plasma beta range in the solar
  wind turbulence.
\newblock {\em Journal of Geophysical Research: Space Physics\/} {\bf 123}
  (2018) 68--75.
\newblock \doi{10.1002/2017JA024813}.

\bibitem[{Franci et~al.(2016)Franci, Landi, Matteini, Verdini, and
  Hellinger}]{Franci2016}
Franci L, Landi S, Matteini L, Verdini A, Hellinger P.
\newblock Plasma beta dependence of the ion-scale spectral break of solar wind
  turbulence: High-resolution 2d hybrid simulations.
\newblock {\em The Astrophysical Journal\/} {\bf 833} (2016) 91.
\newblock \doi{10.3847/1538-4357/833/1/91}.

\bibitem[{Gamayunov et~al.(2015)Gamayunov, Engebretson, Zhang, and
  Rassoul}]{Gamayunov2015}
Gamayunov KV, Engebretson MJ, Zhang M, Rassoul HK.
\newblock Source of seed fluctuations for electromagnetic ion cyclotron waves
  in earth's magnetosphere.
\newblock {\em Advances in Space Research\/} {\bf 55} (2015) 2573--2583.

\bibitem[{Chaston et~al.(2014)Chaston, Bonnell, Wygant, Mozer, Bale, Kersten
  et~al.}]{Chaston2014}
Chaston CC, Bonnell JW, Wygant JR, Mozer F, Bale SD, Kersten K, et~al.
\newblock Observations of kinetic scale field line resonances.
\newblock {\em Geophysical Research Letters\/} {\bf 41} (2014) 209--215.
\newblock \doi{10.1002/2013GL058507}.

\bibitem[{Alexandrova et~al.(2008)Alexandrova, Carbone, Veltri, and
  Sorriso‐Valvo}]{Alexandrova2008}
Alexandrova O, Carbone V, Veltri P, Sorriso‐Valvo L.
\newblock Small‐scale energy cascade of the solar wind turbulence.
\newblock {\em The Astrophysical Journal\/} {\bf 674} (2008) 1153--1157.
\newblock \doi{10.1086/524056}.

\bibitem[{Goldstein et~al.(2015)Goldstein, Wicks, Perri, and
  Sahraoui}]{Goldstein2015}
Goldstein ML, Wicks RT, Perri S, Sahraoui F.
\newblock Kinetic scale turbulence and dissipation in the solar wind: key
  observational results and future outlook.
\newblock {\em Phil. Trans. R. Soc. A.\/} {\bf 373} (2015) 20140147.
\newblock \doi{10.1098/rsta.2014.0147}.

\bibitem[{Franci et~al.(2020)Franci, Sarto, Papini, Giroul, Stawarz, Burgess
  et~al.}]{Franci2020}
Franci L, Sarto DD, Papini E, Giroul A, Stawarz JE, Burgess D, et~al.
\newblock Evidence of a "current-mediated" turbulent regime in space and
  astrophysical plasmas  (2020) arXiv:2010.05048 preprint.

\bibitem[{Franci et~al.(2015)Franci, Landi, Matteini, Verdini, and
  Hellinger}]{Franci2015}
Franci L, Landi S, Matteini L, Verdini A, Hellinger P.
\newblock High-resolution hybrid simulations of kinetic plasma turbulence at
  proton scales.
\newblock {\em The Astrophysical Journal\/} {\bf 812} (2015) 21.
\newblock \doi{10.1088/0004-637x/812/1/21}.

\bibitem[{Matthaeus et~al.(2020)Matthaeus, Yang, Wan, Parashar, Bandyopadhyay,
  Chasapis et~al.}]{Matthaeus2020}
Matthaeus WH, Yang Y, Wan M, Parashar TN, Bandyopadhyay R, Chasapis A, et~al.
\newblock Pathways to dissipation in weakly collisional plasmas.
\newblock {\em The Astrophysical Journal\/} {\bf 891} (2020) 101.
\newblock \doi{10.3847/1538-4357/ab6d6a}.

\bibitem[{Vi{\~n}as et~al.(2015)Vi{\~n}as, Moya, Navarro, Valdivia, Araneda,
  and Mu{\~n}oz}]{vinas2015}
Vi{\~n}as AF, Moya PS, Navarro RE, Valdivia JA, Araneda JA, Mu{\~n}oz V.
\newblock {Electromagnetic fluctuations of the whistler cyclotron and firehose
  instabilities in a Maxwellian and Tsallis-kappa-like plasma}.
\newblock {\em J. Geophys. Res.\/} {\bf 120} (2015).
\newblock \doi{10.1002/2014JA020554}.

\bibitem[{Yoon(2017)}]{Yoon2017}
Yoon PH.
\newblock Kinetic instabilities in the solar wind driven by temperature
  anisotropies.
\newblock {\em Reviews of Modern Plasma Physics\/} {\bf 1} (2017) 4.
\newblock \doi{10.1007/s41614-017-0006-1}.

\bibitem[{Weibel(1959)}]{weibel}
Weibel ES.
\newblock Spontaneously growing transverse waves in a plasma due to an
  anisotropic velocity distribution.
\newblock {\em Phys. Rev. Lett.\/} {\bf 2} (1959) 83--84.
\newblock \doi{10.1103/PhysRevLett.2.83}.

\bibitem[{Sagdeev and Shafranov(1960)}]{sagdeev1960}
Sagdeev R, Shafranov V.
\newblock On the instability of a plasma with an anisotropic distribution of
  velocities in a magnetic field.
\newblock {\em J. Exptl. Theoret. Phys. (U.S.S.R.)\/} {\bf 29} (1960) 181.

\bibitem[{Kennel and Petschek(1966)}]{kennel1966b}
Kennel CF, Petschek HE.
\newblock Limit on stably trapped particle fluxes.
\newblock {\em J. Geophys. Res.\/} {\bf 71} (1966) 1--28.
\newblock \doi{10.1029/JZ071i001p00001}.

\bibitem[{Vi{\~n}as et~al.(1984)Vi{\~n}as, Goldstein, and
  Acu{\~n}a}]{vinas1984}
Vi{\~n}as AF, Goldstein ML, Acu{\~n}a MH.
\newblock Spectral analysis of magnetohydrodynamic fluctuations near
  interplanetary shocks.
\newblock {\em J. Geophys. Res.\/} {\bf 89} (1984) 3762--3774.
\newblock \doi{10.1029/JA089iA06p03762}.

\bibitem[{Brinca and Tsurutani(1987)}]{brinca}
Brinca AL, Tsurutani BT.
\newblock Unusual characteristics of electromagnetic waves exited by cometary
  newborn ions with large perpendicular energies.
\newblock {\em Astron. Astrophys.\/} {\bf 187} (1987) 311--319.

\bibitem[{Yoon(1992)}]{yoon}
Yoon PH.
\newblock Quasilinear evolution of {A}lfv\'en-ion-cyclotron and mirror
  instabilities driven by ion temperature anisotropy.
\newblock {\em Phys. Fluids B\/} {\bf 4} (1992) 3627--3637.

\bibitem[{Moya et~al.(2014)Moya, Navarro, Vi{\~n}as, Mu{\~n}oz, and
  Valdivia}]{moya2014}
Moya PS, Navarro R, Vi{\~n}as AF, Mu{\~n}oz V, Valdivia JA.
\newblock Weak turbulence cascading effects in the acceleration and heating of
  ions in the solar wind.
\newblock {\em Astrophys.\ J.\/} {\bf 781} (2014) 76.

\bibitem[{Hellinger et~al.(2014)Hellinger, Trávníček, Decyk, and
  Schriver}]{hellinger2014}
Hellinger P, Trávníček PM, Decyk VK, Schriver D.
\newblock Oblique electron fire hose instability: Particle-in-cell simulations.
\newblock {\em J. Geophys. Res.\/} {\bf 119} (2014) 59--68.
\newblock \doi{10.1002/2013JA019227}.

\bibitem[{Adrian et~al.(2016)Adrian, Vi{\~n}as, Moya, and Wendel}]{adrian2016}
Adrian ML, Vi{\~n}as AF, Moya PS, Wendel DE.
\newblock Solar wind magnetic fluctuations and electron non-thermal temperature
  anisotropy: Survey of wind-swe-veis observations.
\newblock {\em The Astrophysical Journal\/} {\bf 833} (2016) 49.

\bibitem[{Moya et~al.(2012)Moya, Vi\~{n}as, Mu\~noz, and Valdivia}]{Moya2012}
Moya PS, Vi\~{n}as AF, Mu\~noz V, Valdivia JA.
\newblock Computational and theoretical study of the wave-particle interaction
  of protons and waves.
\newblock {\em Ann. Geophys.\/} {\bf 30} (2012) 1361--1369.
\newblock \doi{10.5194/angeo-30-1361-2012}.

\bibitem[{Dasso et~al.(2005)Dasso, Milano, Matthaeus, and Smith}]{Dasso2005}
Dasso S, Milano LJ, Matthaeus WH, Smith CW.
\newblock Anisotropy in fast and slow solar wind fluctuations.
\newblock {\em The Astrophysical Journal\/} {\bf 635} (2005) L181--L184.
\newblock \doi{10.1086/499559}.

\bibitem[{Horbury et~al.(2005)Horbury, Forman, and Oughton}]{Horbury2005}
Horbury TS, Forman MA, Oughton S.
\newblock Spacecraft observations of solar wind turbulence: an overview.
\newblock {\em Plasma Physics and Controlled Fusion\/} {\bf 47} (2005)
  B703--B717.
\newblock \doi{10.1088/0741-3335/47/12b/s52}.

\bibitem[{Gomberoff et~al.(2004)Gomberoff, Mu{\~n}oz, and
  Valdivia}]{Gomberoff2004}
Gomberoff L, Mu{\~n}oz V, Valdivia JA.
\newblock Ion cyclotron instability triggered by drifting minor ion species:
  Cascade effect and exact results.
\newblock {\em Planet. Space Sci.\/} {\bf 52} (2004) 679--684.

\bibitem[{Navarro et~al.(2020)Navarro, Mu{\~{n}}oz, Valdivia, and
  Moya}]{Navarro2020}
Navarro RE, Mu{\~{n}}oz V, Valdivia JA, Moya PS.
\newblock Feasibility of ion-cyclotron resonant heating in the solar wind.
\newblock {\em The Astrophysical Journal\/} {\bf 898} (2020) L9.
\newblock \doi{10.3847/2041-8213/aba0ae}.

\bibitem[{Gary and Tokar(1985)}]{Gary1985}
Gary SP, Tokar RL.
\newblock {The second-order theory of electromagnetic hot ion beam
  instabilities}.
\newblock {\em J. Geophys.\ Res.\/} {\bf 90} (1985) 65--72.
\newblock \doi{10.1029/JA090iA01p00065}.

\bibitem[{Fried and Conte(1961)}]{zeta}
Fried BD, Conte SD.
\newblock {\em The Plasma Dispersion Function\/} (San Diego, California:
  Academic) (1961).

\bibitem[{Muller(1956)}]{Muller1956}
Muller DE.
\newblock {A Method for Solving Algebraic Equations Using an Automatic
  Computer}.
\newblock {\em Mathematical Tables and Other Aids to Computation\/} {\bf 10}
  (1956) 208--215.
\newblock \doi{10.2307/2001916}.

\bibitem[{Navarro et~al.(2014)Navarro, Moya, Mu{\~{n}}oz, Araneda, Vi{\~{n}}as,
  and Valdivia}]{Navarro2014}
Navarro RE, Moya PS, Mu{\~{n}}oz V, Araneda JA, Vi{\~{n}}as AF, Valdivia JA.
\newblock {Solar wind thermally induced magnetic fluctuations}.
\newblock {\em Phys.~Rev.~Lett.\/} {\bf 112} (2014) 1--5.
\newblock \doi{10.1103/PhysRevLett.112.245001}.

\bibitem[{Moya et~al.(2011)Moya, Mu{\~n}oz, Rogan, and Valdivia}]{Moya2011}
Moya PS, Mu{\~n}oz V, Rogan J, Valdivia JA.
\newblock Study of the cascading effect during the acceleration and heating of
  ions in the solar wind.
\newblock {\em J. Atmos. Solar-Terr. Phys.\/} {\bf 73} (2011) 1390--1397.

\bibitem[{Jian et~al.(2016)Jian, Moya, Vi{\~n}as, and Stevens}]{jian2016}
Jian LK, Moya PS, Vi{\~n}as AF, Stevens M.
\newblock Electromagnetic cyclotron waves in the solar wind: Wind observation
  and wave dispersion analysis.
\newblock {\em AIP Conference Proceedings\/} {\bf 1720} (2016) 040007.
\newblock \doi{10.1063/1.4943818}.

\bibitem[{Wicks et~al.(2016)Wicks, Alexander, Stevens, III, Moya, Viñas
  et~al.}]{wicks2016}
Wicks RT, Alexander RL, Stevens M, III LBW, Moya PS, Viñas A, et~al.
\newblock A proton-cyclotron wave storm generated by unstable proton
  distribution functions in the solar wind.
\newblock {\em The Astrophysical Journal\/} {\bf 819} (2016) 6.

\bibitem[{Krall and Trivelpiece(1986)}]{krall}
Krall NA, Trivelpiece AW.
\newblock {\em Principle of Plasma Physics\/} (San Francisco Press Inc.)
  (1986).

\bibitem[{Stix(1992)}]{stix1992}
Stix T.
\newblock {\em Waves in Plasmas\/} (American Institute of Physics) (1992).

\bibitem[{Yoon et~al.(2012)Yoon, Seough, Kim, and Lee}]{yoon2012b}
Yoon PH, Seough JJ, Kim KH, Lee DH.
\newblock Empirical versus exact numerical quasilinear analysis of
  electromagnetic instabilities driven by temperature anisotropy.
\newblock {\em J. Plasma Phys.\/} {\bf 78} (2012) 47--54.
\newblock \doi{10.1017/S0022377811000407}.

\bibitem[{Seough et~al.(2014)Seough, Yoon, and Hwang}]{seough2014}
Seough J, Yoon PH, Hwang J.
\newblock Quasilinear theory and particle-in-cell simulation of proton
  cyclotron instability.
\newblock {\em Phys. Plasmas\/} {\bf 21} (2014) 062118.
\newblock \doi{http://dx.doi.org/10.1063/1.4885359}.

\bibitem[{Seough et~al.(2015)Seough, Yoon, and Hwang}]{seough2015}
Seough J, Yoon PH, Hwang J.
\newblock Simulation and quasilinear theory of proton firehose instability.
\newblock {\em Phys. Plasmas\/} {\bf 22} (2015) 012303.
\newblock \doi{http://dx.doi.org/10.1063/1.4905230}.

\bibitem[{Galtier and Bhattacharjee(2003)}]{Galtier2003}
Galtier S, Bhattacharjee A.
\newblock Anisotropic weak whistler wave turbulence in electron
  magnetohydrodynamics.
\newblock {\em Physics of Plasmas\/} {\bf 10} (2003) 3065--3076.
\newblock \doi{10.1063/1.1584433}.

\bibitem[{Gary and Smith(2009)}]{Gary2009}
Gary SP, Smith CW.
\newblock Short-wavelength turbulence in the solar wind: Linear theory of
  whistler and kinetic alfv{\'e}n fluctuations.
\newblock {\em Journal of Geophysical Research: Space Physics\/} {\bf 114}
  (2009).
\newblock \doi{10.1029/2009JA014525}.

\bibitem[{Schekochihin et~al.(2009)Schekochihin, Cowley, Dorland, Hammett,
  Howes, Quataert et~al.}]{Schekochihin2009}
Schekochihin AA, Cowley SC, Dorland W, Hammett GW, Howes GG, Quataert E, et~al.
\newblock Astrophysical gyrokinetics: Kinetic and fluid turbulent cascades in
  magnetized weakly collisional plasmas.
\newblock {\em The Astrophysical Journal Supplement Series\/} {\bf 182} (2009)
  310--377.
\newblock \doi{10.1088/0067-0049/182/1/310}.

\bibitem[{Boldyrev et~al.(2013)Boldyrev, Horaites, Xia, and
  Perez}]{Boldyrev2013}
Boldyrev S, Horaites K, Xia Q, Perez JC.
\newblock Toward a theory of astrophysical plasma turbulence at subproton
  scales.
\newblock {\em The Astrophysical Journal\/} {\bf 777} (2013) 41.
\newblock \doi{10.1088/0004-637x/777/1/41}.

\bibitem[{Boldyrev and Perez(2012)}]{Boldyrev2012}
Boldyrev S, Perez JC.
\newblock Spectrum of kinetic-alfv{\'e}n turbulence.
\newblock {\em The Astrophysical Journal\/} {\bf 758} (2012) L44.
\newblock \doi{10.1088/2041-8205/758/2/l44}.

\bibitem[{Arzamasskiy et~al.(2019)Arzamasskiy, Kunz, Chandran, and
  Quataert}]{Arzamasskiy2019}
Arzamasskiy L, Kunz MW, Chandran BDG, Quataert E.
\newblock Hybrid-kinetic simulations of ion heating in alfv{\'e}nic turbulence.
\newblock {\em The Astrophysical Journal\/} {\bf 879} (2019) 53.
\newblock \doi{10.3847/1538-4357/ab20cc}.

\bibitem[{Howes(2015)}]{Howes2015}
Howes GG.
\newblock A dynamical model of plasma turbulence in the solar wind.
\newblock {\em Phil. Trans. R. Soc. A.\/} {\bf 373} (2015) 20140145.
\newblock \doi{10.1098/rsta.2014.0145}.

\bibitem[{Gaelzer et~al.(2015)Gaelzer, Yoon, Kim, and Ziebell}]{Gaelzer2015}
Gaelzer R, Yoon PH, Kim S, Ziebell LF.
\newblock On the dimensionally correct kinetic theory of turbulence for
  parallel propagation.
\newblock {\em Physics of Plasmas\/} {\bf 22} (2015) 032310.
\newblock \doi{10.1063/1.4916054}.

\bibitem[{Kim et~al.(2016)Kim, Yoon, and Choe}]{Kim2016}
Kim S, Yoon PH, Choe GS.
\newblock Spontaneous emission of electromagnetic and electrostatic
  fluctuations in magnetized plasmas: Quasi-parallel modes.
\newblock {\em Physics of Plasmas\/} {\bf 23} (2016) 022111.
\newblock \doi{10.1063/1.4941707}.

\bibitem[{Bowen et~al.(2020)Bowen, Bale, Bonnell, Dudok~de Wit, Goetz, Goodrich
  et~al.}]{Bowen2020}
Bowen TA, Bale SD, Bonnell JW, Dudok~de Wit T, Goetz K, Goodrich K, et~al.
\newblock A merged search-coil and fluxgate magnetometer data product for
  parker solar probe fields.
\newblock {\em Journal of Geophysical Research: Space Physics\/} {\bf 125}
  (2020) e2020JA027813.
\newblock \doi{10.1029/2020JA027813}.

\bibitem[{Hellinger et~al.(2015)Hellinger, Matteini, Landi, Verdini, Franci,
  and Tr{\'a}vn{\'i}cek}]{Hellinger2015}
Hellinger P, Matteini L, Landi S, Verdini A, Franci L, Tr{\'a}vn{\'i}cek PM.
\newblock Plasma turbulence and kinetic instabilities at ion scales in the
  expanding solar wind.
\newblock {\em The Astrophysical Journal\/} {\bf 811} (2015) L32.
\newblock \doi{10.1088/2041-8205/811/2/l32}.

\bibitem[{Yoon and Sarfraz(2017)}]{Yoon2017b}
Yoon PH, Sarfraz M.
\newblock Interplay of electron and proton instabilities in expanding solar
  wind.
\newblock {\em Astrophys. J.\/} {\bf 835} (2017) 246.
\newblock \doi{10.3847/1538-4357/835/2/246}.

\bibitem[{Yan and Lazarian(2011)}]{yan2011}
Yan H, Lazarian A.
\newblock Cosmic ray transport through gyroresonance instability in
  compressible turbulence.
\newblock {\em The Astrophysical Journal\/} {\bf 731} (2011) 35.
\newblock \doi{10.1088/0004-637x/731/1/35}.

\bibitem[{Bruno et~al.(2014)Bruno, Trenchi, and Telloni}]{Bruno2014}
Bruno R, Trenchi L, Telloni D.
\newblock Spectral slope variation at proton scales from fast to slow solar
  wind.
\newblock {\em The Astrophysical Journal\/} {\bf 793} (2014) L15.
\newblock \doi{10.1088/2041-8205/793/1/l15}.

\bibitem[{Martinovi{\'{c}} et~al.(2020)Martinovi{\'{c}}, Klein, Kasper, Case,
  Korreck, Larson et~al.}]{Martinovic2020}
Martinovi{\'{c}} MM, Klein KG, Kasper JC, Case AW, Korreck KE, Larson D, et~al.
\newblock The enhancement of proton stochastic heating in the near-sun solar
  wind.
\newblock {\em The Astrophysical Journal Supplement Series\/} {\bf 246} (2020)
  30.
\newblock \doi{10.3847/1538-4365/ab527f}.

\end{thebibliography}


\end{document}